\documentclass[12pt]{article}
\usepackage[cm]{fullpage}
\usepackage{graphicx,amsmath}

\begin{document}

\begin{center}
{\Large Neutral Pion Production in the Threshold Region} \\[1ex]
Conclusive Symposium of the Collaborative Research Centre 443 \\[2ex]
D. Hornidge\footnote{Email: dhornidge@mta.ca}, \emph{Mount Allison University, Sackville, NB, Canada} \\
A. M. Bernstein, \emph{Massachusetts Institute of Technology, Cambridge, MA,
USA}
\end{center}

\abstract{
We give an overview of the physics motivation and evolution of the neutral
pion photoproduction measurements in the threshold region conducted in the A2
collaboration at MAMI\@. The latest two experiments have been performed with
the almost 4$\pi$ Crystal Ball detector. The first was with a linearly
polarized photon beam and unpolarized liquid-hydrogen target. The data
analysis is now complete and the linearly polarized beam asymmetry along with
differential cross sections provide the most stringent test to date of the
predictions of Chiral Perturbation Theory and its energy region of
convergence.  More recently a measurement was performed using both circularly
polarized photons and a transversely polarized butanol frozen-spin target,
with the goal of extracting both the target and beam-target asymmetries.  From
these we intend to extract $\pi N$ scattering sensitive information for the
first time in photo-pion reactions. This will be used to test isospin
conservation and further test dynamics of chiral symmetry breaking in QCD as
calculated at low energies by Chiral Perturbation Theory.
}

\section{Introduction}

Low-energy pion-nucleon interactions and photo-pion production are of special
interest because the pion, the lightest hadron, is a Nambu-Goldstone Boson
which by its existence represents a clear signature of spontaneous chiral
symmetry breaking in QCD~\cite{Nambu,book}.  The dynamic consequences are that
the production and scattering of low-energy pions are weak in the
s-wave~\cite{W66} and strong in the
p-wave~\cite{book,ChPT-review,ChPT-Baryons}, as is seen clearly in the data
for $\pi N$ scattering and the $\gamma N \rightarrow \pi N$
reaction~\cite{AB-Yangfest}. The physical manifestations include the strong
tensor force in the long range (pion-exchange) part of the nucleon-nucleon
potential~\cite{ChPT-Baryons}.  The s-wave amplitudes are small in $\pi N$
scattering and in the $\gamma^{*} N \rightarrow \pi^0N$ reaction, where
$\gamma^{*}$ is a real or virtual photon. This is true since they vanish in
the chiral limit ($m_u,m_d, m_\pi \rightarrow
0$)~\cite{ChPT-review,ChPT-Baryons}; their small, but non-vanishing values are
measures of explicit chiral symmetry breaking.  Moreover, they are isospin
violating~\cite{W,Martin} since $m_u \neq m_d$~\cite{PDB,Leut}, in addition to
electromagnetic effects. 

The fact that the interactions and production amplitudes are weak at low
energies due to the spontaneous breaking of chiral symmetry in QCD has led to
an effective field theory called Chiral Perturbation Theory
(ChPT)~\cite{ChPT}. Despite its name ChPT is not perturbative in the sense
that in QCD at high energies the coupling constant $\alpha_{s}$ becomes small
and normal perturbation theory is accurate. At low energies, where we are
working, $\alpha_{s}$ becomes large and leads to confinement of the quarks and
gluons so that it is preferable for the effective theory to deal with the
pions and nucleons (and not the quarks and gluons) as the degrees of freedom.
The weak hadron-pion interaction at low energies is what leads to a
perturbative approach (similar to a Taylor series) at low energies; this is
characterized by the small parameters $q/ \Lambda_{\chi}, m/\Lambda_{\chi}$
where $q$ and $m$ are the pion momentum and mass, and $\Lambda_{\chi} \simeq $
1 GeV is the chiral symmetry breaking scale. The lowest order ChPT
calculations reproduced the pre-QCD low-energy theorems~\cite{LET}. The higher
order corrections are defined by a well defined set of counting rules which
govern the forms of the interactions and Feynman diagrams that enter the
calculations at any specified order~\cite{ChPT}. Due to the underlying
spontaneous chiral symmetry breaking, the higher order interactions all
contain derivatives, which makes the p-wave interactions strong leading to the
appearance of the low lying $\Delta$ resonance in the $\pi N$ system. One
recently noticed consequence of the weak s-wave and strong p-wave is the
surprisingly early significance of the contribution of the d-waves in the
$\gamma p \rightarrow \pi^0p$ reaction~\cite{Cesar}. Symmetry imposes a
strict form for each term in the interaction but does not prescribe its
magnitude~\cite{ChPT-review,ChPT}. In practice these parameters (which are
low-energy constants) have been obtained by fitting to data and approximately
agree with model estimates. More recently the low-energy constants have been
obtained with lattice calculations which have provided a striking confirmation
of their values~\cite{lattice-ChPT}. This has been obtained in the purely
mesonic sector (e.g.\ $\pi-\pi$ scattering) for which exact agreement between
experiment~\cite{pi-pi-exp} and theory~\cite{pi-pi-theory} has been obtained.
This agreement required the introduction of a unitary cusp in $\pi-\pi$
scattering due to isospin breaking originating in the isospin breaking
$\pi^0,\pi^{\pm}$ mass difference~\cite{pi-pi-theory} similar in origin to
the unitary cusp in the $\gamma p \rightarrow \pi^0p$ reaction in the
vicinity of the $\gamma p \rightarrow \pi^+n$ threshold~\cite{AB-lq}.  For
the pion-nucleon system, the introduction of the nucleon is an additional
complication for the theory which makes the convergence more difficult and
therefore the calculations are more involved and less accurate than in the
purely mesonic sector.  We are presently at the stage where a careful
experimental comparison of theory and experiment as a function of energy is
required to ascertain the accuracy and the region of validity for this theory
for the $\gamma \pi N$ system.  As has been stressed~\cite{AB-CD06}, any
serious discrepancy between these calculations and experiment must be
carefully examined as a potential violation of QCD; and understanding of QCD
in the non-perturbative region has been considered one of the top ten
challenges in all of physics.

Over an extended period of time the efforts of the A2 collaboration at Mainz
have been focused on accurate measurements of low-energy $\gamma N$ Compton
scattering and pion production reactions to perform tests of the ChPT
predictions. Study of the $\gamma p \rightarrow \pi^0 p$ reaction started with
the original MAMI accelerator and a small detector to observe the $\pi^0
\rightarrow \gamma \gamma$ decay~\cite{Beck}. This followed with increasingly
more accurate experiments to obtain the relatively small cross
section~\cite{Fuchs,Schmidt}. A parallel effort was also carried out at
Saskatoon~\cite{Sask} during this period. The Mainz work has been building up
to the sensitive spin observables~\cite{Schmidt}.  The present generation
photo-pion production experiments that we are concentrating on include
accurate measurements of the cross sections, polarized photon asymmetries, and
polarized target and beam-target asymmetries $T$ and $F$ (defined below).
These experiments have been carried out with circularly and linearly
polarized, tagged photons and with the almost $4\pi$ Crystal Ball and TAPS
detector system.
 
With the exception of pionic atoms, the study of low-energy $\pi N$ scattering
is limited by the fact that pion beams decay, so that experiments below a
kinetic energy of $\simeq$ 20 MeV are generally not feasible~\cite{pi-N}.  We
are pioneering measurements of $\pi N$ scattering at low energies as a final
state interaction in pion photo-production~\cite{AB-lq,AB-IS,AB-review}
through the use of the transverse polarized target asymmetry (time reversal
odd) observable~\cite{obs}.  Photo- and electro-production studies on proton
targets involve the $\pi^0p, \pi^+n$ charge states while conventional
pion-proton interactions are in the $\pi^\pm p, \pi^0n$ states.  This is an
excellent opportunity for testing isospin symmetry which is predicted to be
broken by both electromagnetic and strong interactions due to the mass
difference of the up and down quarks~\cite{W,Martin,PDB,Leut}.  The program
for these measurements has been laid out in a review article on threshold
photo-pion physics~\cite{AB-review}.  These experiments will also provide very
stringent tests of dynamical models~\cite{DMT2001} and predictions based on
chiral symmetry breaking in QCD.

\section{Photon Asymmetry}

\begin{figure}[h!tbp]
\centerline{\includegraphics[width=0.5\textwidth]{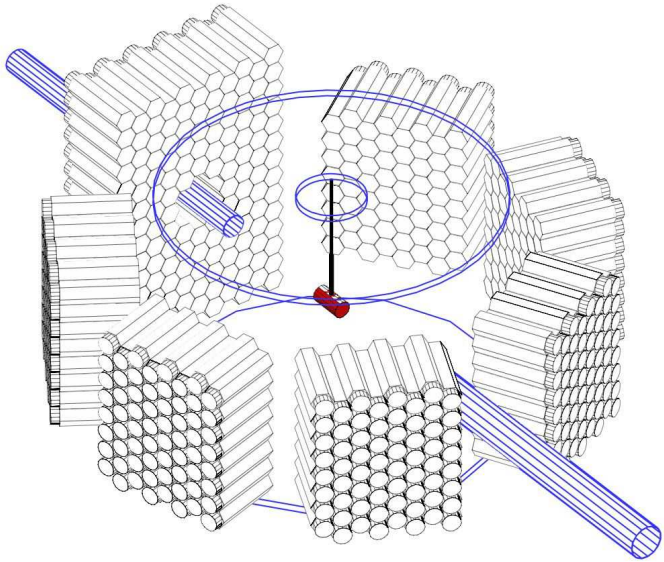}}
\caption{The TAPS detector in the $\pi^0$-detection configuration.  The
solid-angle coverage is approximately 30\% of $4\pi$.}
\label{fig:TAPS}
\end{figure}

\begin{figure}[h!tbp]
\centerline{\includegraphics[width=0.5\textwidth]{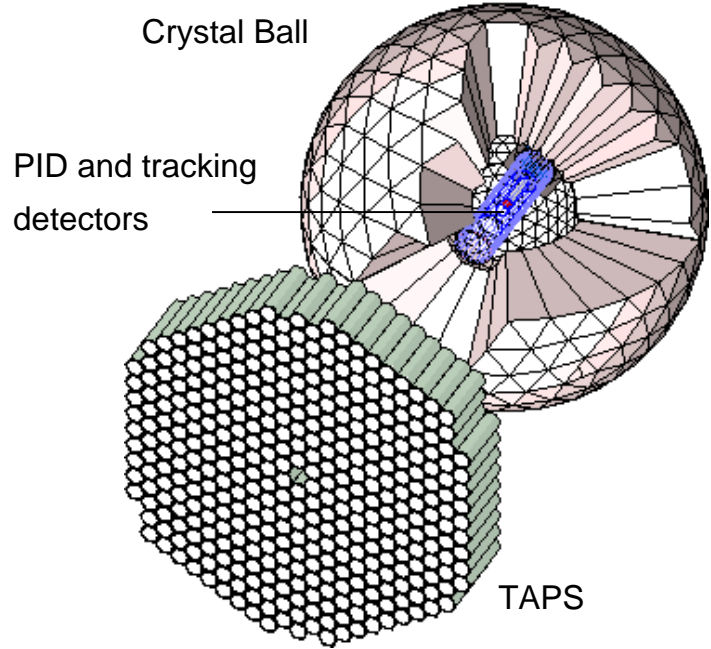}}
\caption{A cut-away view of the CB-TAPS detector system. The solid-angle
coverage is approximately 96\% of $4\pi$.}
\label{fig:CB-TAPS}
\end{figure}

In December 2008 we performed an investigation of the
$\overrightarrow{\gamma}p \rightarrow \pi^0p$ reaction with a linear polarized
photon beam and a liquid H$_2$ target using the Glasgow-Mainz photon tagger
and the CB-TAPS detector system in the A2 hall at MAMI\@.  The purpose was to
perform the most accurate measurement to date of the differential cross
section from threshold through the $\Delta$ region, and to greatly improve our
previous polarized photon asymmetry measurement~\cite{Schmidt}.  The original
experiment was conducted using the TAPS detector alone as shown in
Figure~\ref{fig:TAPS}.
Note that this detector set-up covered only about 30\% of $4\pi$, meaning that
the detection efficiency for the two-photon channel of $\pi^0$ decay was on
the order of 10\%.  The more recent experiment made use of the CB-TAPS set-up
shown in Figure~\ref{fig:CB-TAPS}, which covers $\approx 96\%$ of $4\pi$,
resulting in a detection efficiency for the $\pi^0$ channel of roughly 90\%.
This fact alone made for a large improvement in the accuracy and counting
rates for the new measurement.
In addition, a higher electron beam energy was used which resulted in a
significant increase in the degree of polarization for the incident photon
beam.  The other main difference between the TAPS and CB-TAPS measurements is
that sufficient empty target data was taken for the latter, which turned out
to be crucial due to the contribution to the asymmetry from the $0^+$ nuclei
in the kapton target windows.  Due to poor statistics in the TAPS experiment,
the polarized photon asymmetry, $\Sigma$, was integrated over the entire
incident photon energy range, leading to data only at the cross section
weighted energy average of 159.5 MeV\@.

\begin{figure}[h!tbp]
\centerline{\includegraphics[width=\textwidth]{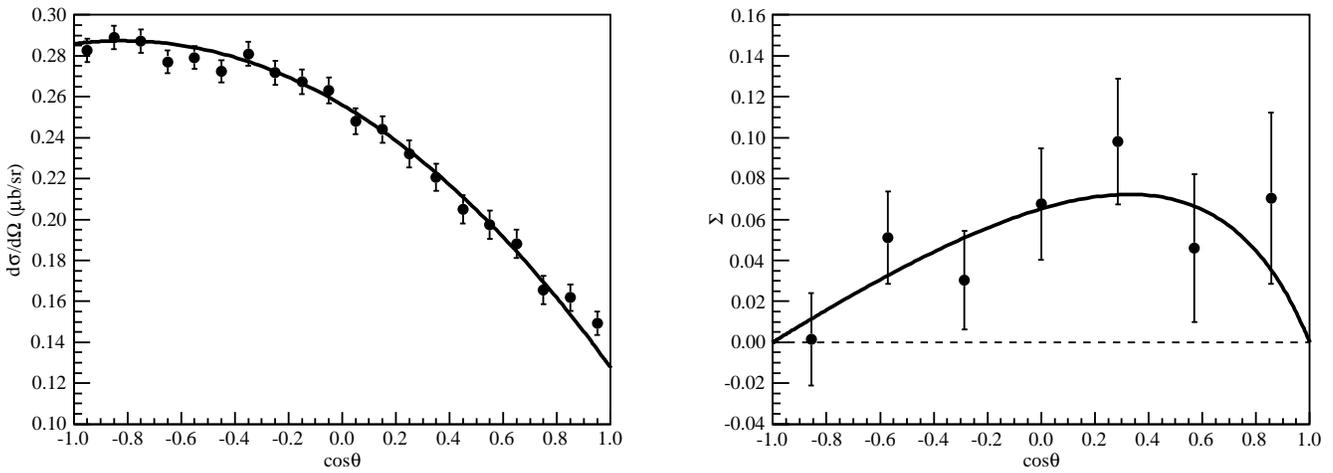}}
\caption{Preliminary CB-TAPS results for a photon energy of 163.9 MeV\@.  Left Panel:
Differential cross section versus $\cos\theta$.  Right panel: Photon asymmetry
versus $\cos\theta$.  The errors are statistical and the lines are preliminary
ChPT fits with s-, p-, and d-waves~\cite{Cesar}.}
\label{fig:Sig}
\end{figure}

\begin{figure}[h!tbp]
\centerline{\includegraphics[width=\textwidth]{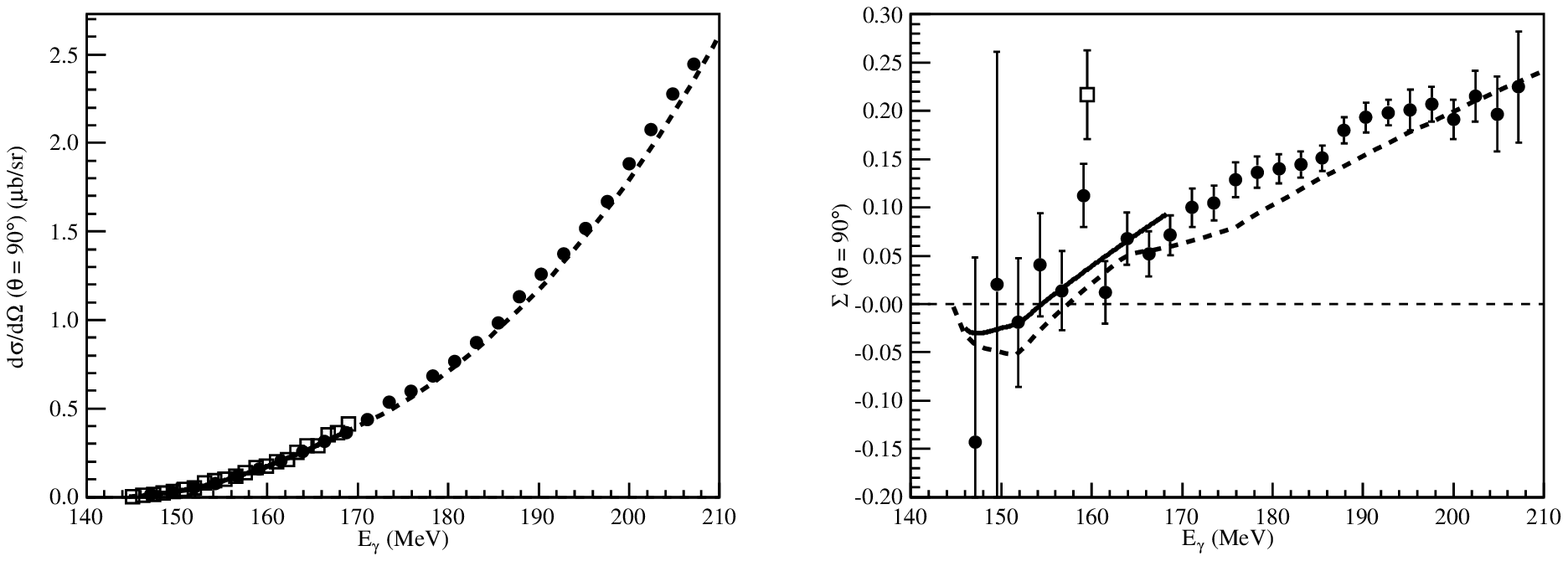}}
\caption{Preliminary CB-TAPS results (solid circles) of the differential cross
section and photon asymmetry at pion CM angle of 90$^\circ$ as a function of
incident photon energy compared to the older TAPS data~\cite{Schmidt} (open
squares) as well as theory.  The solid lines are preliminary ChPT fits to the
new data~\cite{Cesar} and the dashed lines are a dynamical
model~\cite{DMT2001}.  The ChPT fits have been done up to 165 MeV, but the
issue of maximum energy of convergence will be explored.  Errors are
statistical only.  Left Panel: Differential cross section.  Right panel:
Photon asymmetry.}
\label{fig:Sig2}
\end{figure}

The data analysis is close to finished and the results for the differential
cross section and photon asymmetry are shown in Figure~\ref{fig:Sig} at one
photon energy (163.9 MeV) to give an idea of the accuracy.  We have have
photon asymmetries from just above threshold in 
2.4-MeV-wide bins, and differential cross sections from threshold into the
$\Delta$ region.  Fitting of the data has commenced for the low-energy
constants in ChPT~\cite{loop} in collaboration with C.~Fern\'{a}ndez
Ram\'{i}rez.  The solid curves in Figure~\ref{fig:Sig} are the ChPT
calculations using s-, p-, and d-waves~\cite{Cesar} with the low-energy
parameters fit to the data.
A comparison of the new CB-TAPS data with the original TAPS measurement is
given in Figure~\ref{fig:Sig2} along with the ChPT calculations with updated
low-energy parameters and the 2001 version of the DMT dynamical
model~\cite{DMT2001}.

With the use of a model-independent partial-wave analysis, one can extract
various coefficients from the differential cross sections and photon beam
asymmetry, and then comparison can be made between the extracted coefficients
and the theory predictions.  The s-, p-, and d-wave multipoles then appear
only in the coefficients which allows for a very direct comparison of theory
and experiment.  In particular, the differential cross section can be expanded
in terms of the pion CM angle, $\theta$, in the following way
\begin{equation*}
\frac{d\sigma}{d\Omega}(\theta) = A_T + B_T\cos\theta + C_T\cos^2\theta
\end{equation*}
where $A_T$, $B_T$, and $C_T$ are the coefficients.  The photon beam asymmetry
is related to the transverse-transverse cross section
\begin{equation*}
\frac{d\sigma_{TT}}{d\Omega}(\theta) = \sin^2\theta(A_{TT} + B_{TT}\cos\theta
	+ C_{TT}\cos^2\theta)
\end{equation*}
through the polarized photon asymmetry $\Sigma$, where 
\begin{equation*}
\Sigma(\theta) = -\frac{d\sigma_{TT}}{d\Omega}(\theta)/ \frac{d\sigma}{d\Omega}
(\theta).
\end{equation*}
Here the effects of the d-waves will appear in all coefficients, but it is the
$B_{TT}$ coefficient where the effect is the most dramatic since it equals 0
if only s- and p-waves contribute.  Our preliminary analysis indicates
significant non-zero values which is the first direct experimental proof that
the d-waves do contribute at low energies as predicted~\cite{Cesar}. 

Analysis of the coefficients and multipoles in currently ongoing, and once
finished will allow an accurate extraction of the s- and all three p-waves.
More important, for the first time, the energy dependence of the p-waves will
be obtained along with a definitive determination of d-wave contributions.
The unitary cusp in the s-wave amplitude arising from charged pion
re-scattering will also be examined, leading to the extraction of the cusp
function for the real part of the electric dipole amplitude.  These data
provide the most stringent test to date of the predictions of Chiral
Perturbation Theory and its energy region of convergence.

\section{Target and Beam-Target Asymmetries}

\begin{figure}[h!tbp]
\centerline{\includegraphics[width=\textwidth]{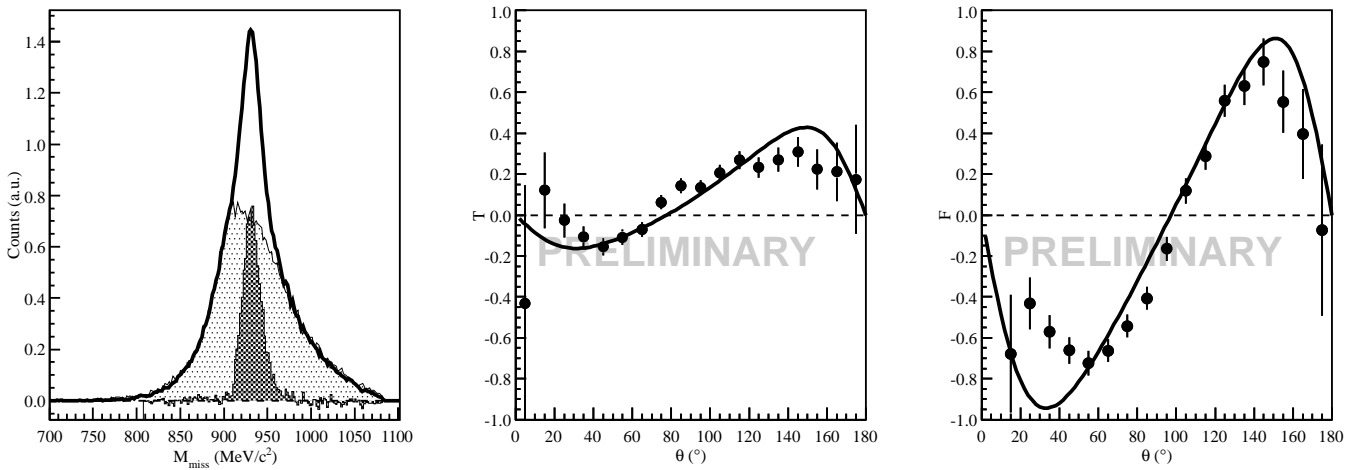}}
\caption{Preliminary results for a photon energy of 320 MeV\@.  Left panel:
Counts versus missing mass for the butanol target (unfilled histogram), carbon
target (light grey histogram), and the difference due to the protons in the
target (dark grey histogram).  Centre panel: $T$ asymmetry versus $\theta$
(the pion CM angle).  Right panel: $F$ asymmetry versus $\theta$.  The errors
are statistical and the lines are predictions of the MAID
model~\cite{MAID}.}
\label{fig:FT}
\end{figure}

The development of a transverse polarized proton target (butanol frozen spin)
has enabled us to access time reversal odd observables which are sensitive to
the phases of the $\pi N$ final states~\cite{AB-lq,AB-IS,AB-review,Anant}.  We
have performed precise measurements of the
$\overrightarrow{\gamma}\overrightarrow{p} \to \pi^0p,\pi^+p$ reactions from
threshold to the $\Delta$ resonance using circularly polarized photon beams
and a transverse polarized target~\cite{A2-prop}.  We have measured the
polarized target asymmetry $\mathbf{T=A(y)}$~\cite{AB-review,obs} for which
the target polarization is perpendicular to the reaction plane.  We also
measured the double polarization observable $F=A(\gamma_c,x)$ (circular
polarized photons on a transverse polarized target)~\cite{AB-review,obs} which
is sensitive to the d-wave multipoles that have recently been shown to be
important in the near threshold region~\cite{Cesar}.  Preliminary data from
this experiment is shown in Figure~\ref{fig:FT} for a photon energy of 320
MeV\@.  The first panel shows the missing mass plot which is useful to
separate the $\pi^0$ mesons produced from the protons from the other elements
in the butanol target and in the target cell walls.  We also have performed a
background experiment in which a foamy carbon target with the same geometry as
the butanol was measured.  By subtracting the suitably normalized carbon
target data from that of the butanol it can be seen that the proton target
signal can be accurately extracted.  Using this technique a preliminary
analysis of part of the data are presented for the $T$ and $F$ asymmetries and
compared to the predictions of the MAID model.  Analysis of the data is
currently under way.

The $T$ asymmetry (time reversal odd) will measure the charge exchange
scattering length $a_{cex}(\pi^+n \rightarrow \pi^0p)$ from the unitary cusp
above the $\pi^+n$ threshold~\cite{AB-lq,AB-review}, which is a measure of the
s-wave interaction between two unstable particles! We anticipate $\simeq$ 1\%
statistical and $\leq$ 2\% systematic uncertainties, where the latter is
dominated by the degree of target polarization.  If isospin is conserved then
$a_{cex}(\pi^+n \rightarrow \pi^0p) = a_{cex}(\pi^-p \rightarrow \pi^0n)$.  At
the present time the right-hand side has been measured in pionic hydrogen with
an error of $\simeq$ 1.5\%~\cite{Gotta}, and it is anticipated that future
work will reduce the uncertainty.  Any deviations from the isospin conserving
limit will test isospin breaking due to the electromagnetic interaction and
the strong interaction due to the mass difference of the up and down quarks
predicted in ChPT~\cite{Martin}.  Observation of $T$ for the first time in the
intermediate-energy region, combined with the other accurate data which we are
obtaining, will provide us with information about the $\pi N$ phase shifts for
charge states ($\pi^0p,\pi^+n$) that are not accessible to conventional $\pi
N$ scattering experiments.  This will enable us to test isospin
conservation~\cite{AB-IS}.  In addition these measurements will test detailed
predictions of Chiral Perturbation Theory~\cite{loop} and its energy region of
convergence.

\end{document}